\documentclass[aps,pra,superscriptaddress,twocolumn,footinbib,showpacs,floatfix]{revtex4-1}

\usepackage{graphicx}% Include figure files
\usepackage{dcolumn}% Align table columns on decimal point
\usepackage{bm}% bold math
\usepackage{color}
\usepackage[colorlinks=true,urlcolor=blue,citecolor=blue, linkcolor = blue]{hyperref}
\usepackage{amssymb,ulem,amsmath}
\def\be{\begin{equation}}
\def\ee{\end{equation}}
\def\ba{\begin{array}{lll}}
\def\ea{\end{array}}
\def\ber{\begin{eqnarray}}
\def\eer{\end{eqnarray}}

\begin{document}
\title{Phase-induced transport in atomic gases: from superfluid to Mott insulator}
%\title{Crossover from continuum to Mott insulator in phase induced transport}
%
\author{Sebastiano Peotta}
\email{speotta@physics.ucsd.edu}
\affiliation{Department of Physics, University of California, San Diego, California 92093, USA}
\author{Chih-Chun Chien}
\email{chienchihchun@gmail.com}
\affiliation{School of Natural Sciences, University of California, Merced, CA 95343, USA}
\author{Massimiliano Di Ventra}
\email{diventra@physics.ucsd.edu}
\affiliation{Department of Physics, University of California, San Diego, California 92093, USA}
\begin{abstract}
Recent experimental realizations of artificial gauge fields for cold atoms are promising for generating steady states carrying a mass current in strongly correlated systems, such as the Bose-Hubbard model. Moreover, a homogeneous condensate confined by hard-wall potentials from laser sheets has been demonstrated, which provides opportunities for probing the intrinsic transport properties of isolated quantum systems. Using the time-dependent Density Matrix Renormalization Group (TDMRG), we analyze the effect of the lattice and interaction strength on the current generated by a quench in the artificial vector potential when the density varies from low values (continuum limit) up to integer filling in the Mott-insulator regime. There is no observable mass current deep in the Mott-insulator state as one may expect. Other observable quantities used to characterize the quasi-steady state in the bulk of the system are the Drude weight and entanglement entropy production rate. The latter in particular provides a striking signature of the superfluid-Mott insulator transition. 
Furthermore, an interesting property of the superfluid state is the formation of shock and rarefaction waves at the boundaries due to the hard-wall confining potentials. We provide results for the height and the speed of the shock front that propagates from the boundary toward the center of the lattice. Our results should be verifiable with current experimental capabilities.
%The result of the quasi-exact TDMRG method are compared to that from a mean-field time-dependent Gutzwiller ansatz. The current as a function of filling and interaction strength is well captured by the Gutzwiller ansatz except close to the superfluid-insulator transition. 
\end{abstract}
\maketitle
\section{Introduction\label{sec:intro}}
Ultracold atomic gases are among the most successful implementations of a quantum simulator~\cite{Cirac:2012,Nori:2014}. Some paradigms in condensed matter physics have analogues in ultracold atomic gases and can be studied in an ideal setting with full control of the Hamiltonian parameters. For instance, the microcanonical approach to quantum transport~\cite{DiVentra:2004,Bushong:2005,Chien:2014} that has been used to test certain assumptions of the scattering approach to conduction in nanoscale systems, can now be fully realized in cold-atom systems with relative ease, thus providing a direct test of several predictions that are difficult to verify in the solid state~\cite{Chih-Chun:2012,ChienEPL:2012,ChienNJP:2013,Chien:2014}.

Recent advances in experiments have demonstrated artificial electric and magnetic fields from artificial gauge fields for cold atoms \cite{Lin:2009a,Lin:2009b,Lin:2009c,Lin:2011,Dalibard:2011}, which offer the opportunity to study a great variety of problems relevant to conventional condensed matter systems. When cold atoms are confined in optical lattices, the hopping coefficients can acquire a phase via Peierls substitution \cite{Hofstadter:1976} using artificial gauge fields \cite{PS1:2012} or lattice modulations \cite{PS2:2012}. For example, charge and spin transport in strongly correlated systems~\cite{Raizen:1997,Ott:2004,Strohmaier:2007,Cramer:2008,Hung:2010,Heidrich-Meisner:2010,Bruderer:2012,Schneider:2012,Brantut:2012,Beeler:2013,Chih-Chun:2013,Beeler:2013,Ronzheimer:2013,Vidmar:2013,Carrasquilla:2013} are among the most interesting problems that can now  be addressed from a different perspective using ultracold atomic gases driven by artificial gauge fields.

\begin{figure}
\includegraphics{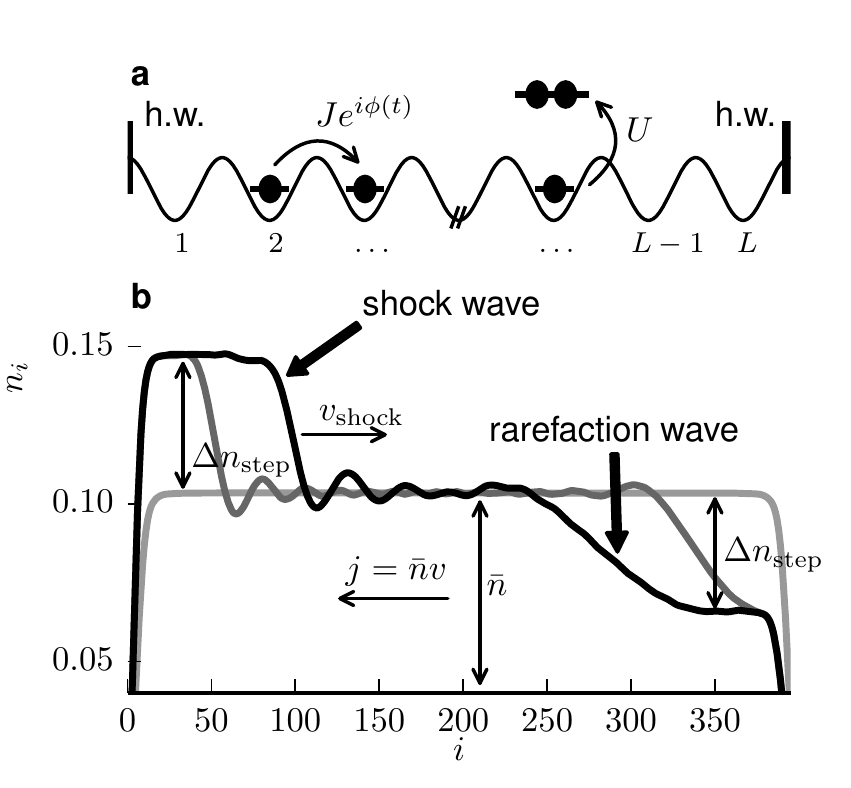}
\caption{\label{fig:one_1} \textbf{a)} Schematic representation of the Bose-Hubbard Hamiltonian with a complex time-dependent hopping term $Je^{i\phi(t)}\hat{b}^\dagger_i\hat{b}_{i+1} + \mathrm{H.c.}$ and interaction term $U\hat{n}_i(\hat{n}_i-1)/2$. The system is confined by hard-wall (h.w.) boundaries at the edges of the lattice~\cite{Hadzibabic:2013,Hadzibabic:2014,Hadzibabic:2014-2}. \textbf{b)} Density profiles observed after the quench of the hopping phase $\phi(t <0) = 0 \to \phi(t \geq 0) = 0.05$ for $U/J = 0.4$. The light grey profile corresponds to $t=0$ ($t_0 = \hbar/J$), the grey to $t=200\,t_0$ and the black to $t= 400\, t_0$. The bulk current $j = \bar{n}v$ is driven to the left of the system by the phase quench ($\bar{n}$ is the bulk density in the initial state and $v$ the particle velocity). A shock wave with height $\Delta n_\text{step}$ and front speed $v_\text{shock}$ forms at the left boundary, while a rarefaction wave with the same height forms at the right boundary.}
\end{figure}

In this regard, the superfluid/Mott-insulator transition~\cite{Fisher:1989,Buchler:2003} in the Bose-Hubbard model has been realized in cold-atom systems \cite{Greiner:2002} and subsequently studied in a large number of papers (see \cite{BHreview} for a review). Here, we investigate transport properties of the Bose-Hubbard model by means of a sudden change of the hopping phase that delivers a finite momentum to the gas (see Fig.~\ref{fig:one_1}\textbf{a}). For low filling or weak interactions the system is close to the continuum limit and the atoms are delocalized. For integer filling and strong enough interactions the atoms localize and the system becomes a Mott insulator. It is important to address the issue of how the lattice-induced correlations affect the transport in between these two limits, as the system is tuned from the weakly-interacting regime to the strongly interacting one.

In order to achieve this goal, we employ the quasi-exact density matrix renormalization group (DMRG) method using a matrix product state (MPS) ansatz for the wavefunction. Recently the static DMRG method has been applied to the study of the Bose-Hubbard model phase diagram under the influence of artificial gauge fields~\cite{Zhao:2014,Piraud:2014,Xu:2014,Peotta:2014b}.
As a step further we study the evolution in time using the time-dependent DMRG algorithm (TDMRG)~\cite{White:2004,Vidal:2004,Daley:2004,tdmrg} within the microcanonical picture of transport~\cite{DiVentra:2004,Bushong:2005,Chien:2014}, which is ideal to study transport phenomena in closed finite systems as the present ones. 

We concentrate on the case of ultracold bosons in a one-dimensional optical lattice with a superimposed external confining potential which is constant throughout the system and rises sharply at the boundary. Hard-wall confining potentials of this kind are not common in ultracold atoms where harmonic confining potentials are the norm, but have been recently realized~\cite{Hadzibabic:2013}, and the ground state properties of uniform condensates have been measured~\cite{Hadzibabic:2014,Hadzibabic:2014-2}. They offer the advantage that in a uniform system one can focus on the intrinsic transport properties without spurious effects due to the external confinement. Moreover, boundary effects such as the density waves studied here may not be observable in a harmonic confinement where the whole cloud can move together in a sloshing fashion.

After a sudden quench of the hopping phase to a non-zero value, the gas in the superfluid phase is quickly driven into a quasi-steady state with constant current that does not decay with time. We use the quasi-exact TDMRG to extract the quasi-steady state current as a function of filling and interaction strength and we argue that in fact this corresponds to measuring the Drude weight~\cite{Kohn:1964}. Ultimately this technique can be used to infer the Drude weight or, possibly, the superfluid fraction in real experiments in a simple way.
Near the superfluid-Mott insulator transition the current decays quickly from the finite value attained immediately after the quench, as expected from the vanishing Drude weight.

%%%%%%%%%%%%%%%%%%
%%%%% FIG. 1 %%%%%
%%%%%%%%%%%%%%%%%%

Another important quantity characterizing the quasi-steady state is the entanglement entropy between two parts of the system. Recently it has been found that the rate at which entanglement entropy is generated between two Fermi seas is related to the noise statistics of the current through a quantum point contact that couples the two reservoirs~\cite{KlichLevitov}. Since fluctuations can be accessed in the ultracold gases counterpart of solid state transport experiments~\cite{Jacqmin:2011,Klawunn:2011,Armijo:2012}, we report results for the entanglement entropy production rate in the Bose-Hubbard model driven by a quench in the hopping phase and we find that the entropy production rate is non-zero only around the superfluid-Mott insulator transition. This can serve as an alternative way to detect the transition in experiments. 

The quasi-steady state is not expected to persist indefinitely in a system with hard-wall boundary conditions. In our case we observe that shock and rarefaction waves form at the boundaries and propagate toward the center of the chain as illustrated in Fig.~\ref{fig:one_1}\textbf{b}. We assume that the continuously shrinking middle region in between the two waves is a good approximation of a true steady state, since the current reaches its stationary value on a short time scale and is independent of the position within the same region. Moreover, in recent experiments~\cite{Cheneau:2012} it has been observed that perturbations (in our case the boundary effects) propagate at a finite speed, in a light cone manner. It is natural to assume in our case that the speed of propagation of the perturbations is in fact the shock front speed $v_\mathrm{shock}$ (see Fig.~\ref{fig:one_1}\textbf{b}).

The quench that we consider here is in essence the \textit{piston problem} of shock wave theory for a fluid in a lattice rather than in the continuum. In the piston problem a piston is moved at constant speed in a fluid initially at rest in a long cylinder. The piston problem for dispersive shock waves has been considered recently for a fluid described by the nonlinear Schr{\"o}dinger equation~\cite{Hoefer:2008}. We find that, regardless of the presence of the lattice, dispersive shock and rarefaction waves form whenever the system is a superfluid, but not in the Mott insulator. From our simulations we extract the height of the well defined density step $\Delta n_\mathrm{step}$ that precedes the shock wave or, equivalently, the wave speed $v_\mathrm{shock}$ (see Fig.~\ref{fig:one_1}\textbf{b}). We conclude that dispersive waves are a general signature of the superfluid.
We emphasize that the formation of such waves is a consequence of the hard wall potentials at the boundaries. If  the system is confined in a harmonic trap, the same kind of quench would simply trigger a periodic oscillation of the center of mass, eventually damped by lattice effects~\cite{Fertig:2005,Rey:2005,Montangero:2009}, but waves do not form.

Note that recent works have focused on the dynamics of Fermi- and Bose-Hubbard models under a constant force~\cite{Heidrich-Meisner:2010,Carrasquilla:2013}. The quench protocol considered here is distinct and leads to a quite different dynamics. The difference with respect to the case of an applied constant force for $t>0$ is that the quench considered here corresponds to a large force applied for a short time (impulsive force), after which the system evolves freely. A constant force applied to a lattice system generally leads to Bloch oscillations of the gas center of mass~\cite{Raizen:1997,Heidrich-Meisner:2010}, but no such oscillations have been observed in our case. On the other hand, a large tilted lattice potential applied for a small time can be used to approximate the phase quench considered here. This offers an alternative way to experimentally verify our results.

The paper is organized as follows. Section~\ref{sec:BH} briefly reviews the Bose-Hubbard model and provides details on the TDMRG method used to study it. In Section~\ref{sec:noninteracting} the limiting cases of noninteracting bosons and fermions are discussed, as well as the mapping of the Bose-Hubbard model into the XXZ spin chain. In Section~\ref{sec:dynamics} the bulk properties (away from the boundaries) of the quasi-steady state are considered, namely current (Sec.~\ref{sec:steady}), Drude weight (Sec.~\ref{sec:drude}) and entanglement entropy (Sec.~\ref{sec:entanglement}). In Section~\ref{sec:shock} is focused on the density profile dynamics at the boundaries which is interpreted as the formation and propagation of shock and rarefaction waves. Finally, Section~\ref{sec:conclusion} summarizes the main results.
%We discuss first the results for the small phase quench (Sections~\ref{sec:shock},~\ref{sec:steady},~\ref{sec:drude} and~\ref{sec:wavefront}) since in this case the TDMRG data are available for longer times. We will comment at the end the data for the large phase quench (Section~\ref{sec:large_phase}).

\section{Bose-Hubbard model and TDMRG simulations}\label{sec:BH}
In the presence of a vector potential, the Bose-Hubbard Hamiltonian reads (Fig.~\ref{fig:one_1}\textbf{a})
\begin{equation}\label{eq:ham}
\begin{split}
\mathcal{\hat H}_{\rm BH} = &-J \sum_{i = 1}^{L-1}(e^{i\phi(t)}\hat b_{i}^\dagger \hat b_{i+1}+\mathrm{H.c.}) \\ &+ \frac{U}{2}\sum_{i=1}^L\hat n_i(\hat n_i-1) + \sum_i V_i {\hat n}_i\,,
\end{split}
\end{equation}
where $\hat b_i$, $\hat b_i^\dagger$ are bosonic annihilation and creation operator respectively, satisfying the canonical commutation relations $[\hat b_i;\hat b_i^\dagger] = 1$, and $\hat n_i = \hat b^\dagger_i\hat b_i$ are the corresponding number operators. The hopping coefficient acquires a phase $\phi(t)=\int \mathbf{A}\cdot d\mathbf{l}$ via Peierls substitution \cite{Hofstadter:1976}, where $\mathbf{A}$ is the vector potential. Throughout the paper time is measured in units of $t_0\equiv \hbar/J$ and and energies in unit of $J$. The lattice constant $a$ is taken as the unit of length, thus the current and velocities are measured in units of $t_0^{-1}$.

In order to apply the TDMRG method we will use finite lattices of different lengths $L$. The filling $n=N/L$ is controlled by the number of particles $N$ in the lattice.
We report the results for the following selected values of the filling: $n = 0.1$ ($L = 400,\,N= 40$), $n = 0.25$ ($L = 160,\,N= 40$), $n = 0.5,\,0.75,\,1.0$ ($L = 100,\,N= 50,\,75,\,100$). The gas is confined only by the hard-wall boundaries and the external potential in the bulk of the system is uniform ($V_i = 0$ in Eq.~\eqref{eq:ham} and~\eqref{eq:XXZ}). Experimentally this has been realized using two sheet laser beams that create sharp repulsive potentials at the boundaries~\cite{Hadzibabic:2013,Hadzibabic:2014,Hadzibabic:2014-2}. 

The system is initially prepared in the ground state of the Hamiltonian~(\ref{eq:ham}) without any vector potential [$\phi(t< 0) = 0$]. An artificial vector potential is suddenly applied to the system so that the hopping coefficient acquires a finite phase $\phi(t\geq 0) = \phi_0$. A phase quench amounts to a rearrangement of energy eigenstates so the system is driven out of equilibrium. We consider two values for the post-quench phase: a small value $\phi_0 = 0.05$ and a large one $\phi_0=0.5$.  

In the TDMRG simulations the link dimension $m$ of the MPS matrices is adjusted automatically in time and space by requiring a fixed truncation error of  $\varepsilon = 10^{-10}$~\cite{tdmrg}. However, $m$ is not allowed to be larger than $m =100$ during the ground state optimization and $m =300$ during the dynamics in the case of the small phase quench. A larger $m$ is used during the dynamics in order accommodate the entanglement generated as the system is driven out of equilibrium. We use the upper limits $m = 500$ and $m=2000$ for the static and dynamic DMRG respectively, in the case of the large phase quench since the entanglement is generally larger for a system that is driven farther out of equilibrium. Occasionally we observed that the  required truncation error is not always met during the evolution because of the upper limit on $m$. However, we have verified that the local observables that we are interested in, namely the density, current and entanglement entropy, are only slightly affected by the larger truncation error and the level of precision provided by the above parameters is sufficient for our purposes.

The TDMRG results are also compared to a  time-dependent mean-field approximation based on the \textit{grand-canonical} Gutzwiller ansatz~\cite{Jaksch:2002,Kollath:2012}, \textit{i.e.}, an MPS with \textit{link dimension} $m = 1$. The Gutzwiller wavefunction has been evolved in time with a variation of the TDMRG algorithm as explained in Ref.~\cite{Peotta:2013}, and we find a remarkably close agreement away from the Mott insulator state, a useful information since the the Gutzwiller ansatz is much less computationally demanding than full TDMRG simulations.

\section{Noninteracting and hardcore bosons}\label{sec:noninteracting}
Before analysing the TDMRG results we briefly discuss the noninteracting limits of the Hamiltonian~(\ref{eq:ham}) focusing first on the transition between noninteracting ($U = 0$) and interacting ($U>0$) bosons.
The ground state of noninteracting bosons with a fixed number of particles shows different features when compared to that of bosons with finite interactions $U$~\cite{Giamarchi_book}. The former is a condensate with all the particles occupying the lowest available state, which in the case of a box potential in 1D has a density profile $n_j \propto \sin^2\left(\frac{\pi j}{L+1}\right)$, while for bosons with a finite $U$ the density profile is flat in a finite region in the middle of the system for large enough $L$.

In fact, a phase transition occurs when $U$ is changed from exactly zero to any finite value. To see this explicitly we note that for small values of the interaction strength the Bose-Hubbard model is well approximated by a continuum field theory of bosons with delta function interaction $g_B\delta(x_1-x_2)$~\cite{Delande:2014} known as the Lieb-Liniger model~\cite{Lieb:1963a,Lieb:1963b}. In the following we will frequently use the density-rescaled interaction strenth $\gamma = U/(2Jn)$ which coincides with the Lieb-Liniger parameter $\gamma = mg/(\hbar^2 \rho)$~\cite{Lieb:1963a} ($m$ is the particle mass and $\rho$ the density in the continuum). A gas can be considered weakly interacting if $\gamma < 1$ while it is strongly interacting if $\gamma >1$. The Bose-Fermi mapping valid for \textit{arbitrary} values of $g_B$ ensures that bosons in the continuum are equivalent to fermions with a $p$-wave interaction $g_F\delta'(x_1-x_2)$ where $g_F = -4(\hbar^2/m)^2/g_B$ (with $m$ the particle mass)~\cite{Cheon:1998,Cheon:1999}. Discretizing the fermionic Hamiltonian results in~\cite{Muth:2011}
\begin{equation}\label{eq:XXZ}
\begin{split}
\mathcal{\hat H}_{\rm XXZ} = &-J\sum_i(e^{i\phi(t)}\hat c_i^\dagger\hat c_{i+1} + {\rm H.c.}) \\&-\frac{2J}{1+U/(4J)}\sum_i\hat n_i\hat n_{i+1} +\sum_iV_i\hat n_i\,,
\end{split}
\end{equation}
with fermionic annihilation and creation operators $\hat c_i$,$\hat c_i^\dagger$ and density operator $\hat n_i = \hat c_i^\dagger\hat c_i$.

%%%%%%%%%%%%%%%%%%
%%%%% FIG. 2 %%%%%
%%%%%%%%%%%%%%%%%%
\begin{figure*}
\includegraphics{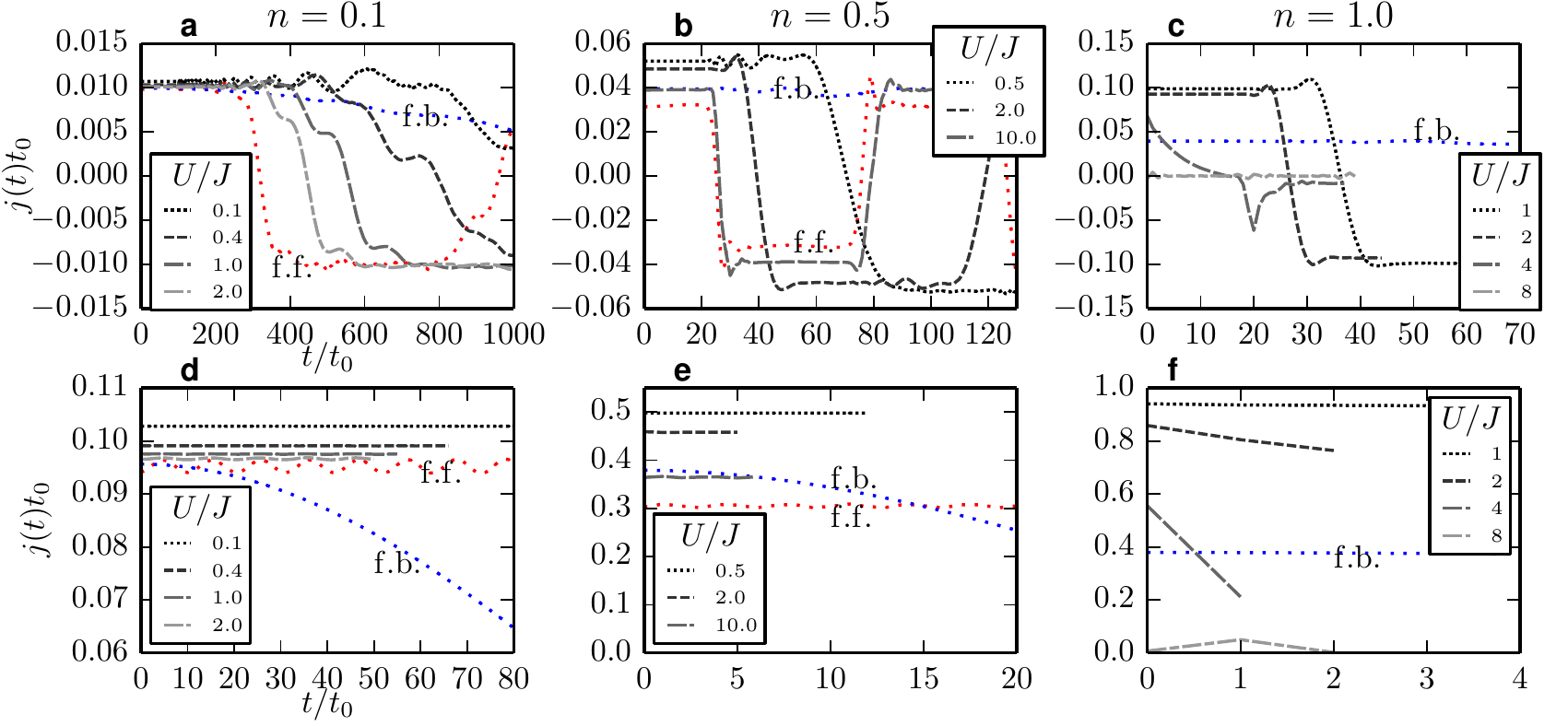}
\caption{\label{fig:two} (Color online) \textbf{Panels} \textbf{a}-\textbf{b}-\textbf{c}: current $j(t)$ in the middle of the chain [Eq.~(\ref{eq:current_exp})] as a function of time $t$ for the fillings $n = 0.1$~(panel \textbf{a}), $n = 0.5$~(\textbf{b}) and $n = 1$~(\textbf{c}) and selected values of the Hubbard interaction $U/J$ after quenching the hopping phase to the value $\phi_0 = 0.05$ (small phase quench). The blue and red dotted lines are the results for free bosons (f.b., $U = 0$) and hardcore bosons, equivalent to free fermions (f.f., $U = +\infty$), respectively.
\textbf{Panels} \textbf{d}-\textbf{e}-\textbf{f}: same as the above panels after a quench of the hopping phase to the value $\phi_0 = 0.5$ (large phase quench). The filling are the same as in the panels immediately above. Note the non-stationary character of the current induced by the quench in the case of free bosons.}
\end{figure*}

The above fermionic Hamiltonian can be readily recasted into the XXZ spin model~\cite{Korepin_book} by a Jordan-Wigner transformation~\cite{Lieb:1961}. The XXZ model is gapped when the anisotropy parameter
\begin{equation}\label{eq:anisotropy_parameter}
\Delta=-\frac{1}{1+U/(4J)}
\end{equation}
satisfies $|\Delta|\geq 1$ and it is gapless otherwise. The transition point from noninteracting to interacting bosons in the original Bose-Hubbard model corresponds to $\Delta =-1$, the ferromagnetic transition of the XXZ model. This phase transition is reflected in the absence of a stationary steady state with finite current in the thermodynamic limit of noninteracting bosons~\cite{Chih-Chun:2013}.
%, which is related to the inhomogeneity of the density within any finite segment of the system. This implies that the boundaries always play a significant role in the dynamics of noninteracting bosons, no matter how large the system is. 
As shown in the following this is particularly evident in the case of a large phase quench.

The fermionic Hamiltonian in Eq.~(\ref{eq:XXZ}) is equivalent to the Bose-Hubbard model (\ref{eq:ham}) only in the low filling limit since the Bose-Fermi mapping for 1D particles in the continuum has been implemented as an intermediate step. This is evident from the fact that Eq.~(\ref{eq:XXZ}) is integrable for any value of $U/J$ while the Hamiltonian (\ref{eq:ham}) is not. However, in the hardcore-boson limit $U/J\to +\infty$ (corresponding to free fermions, $\Delta= 0$) the two are equivalent. Below we will compare the dynamics for finite $U/J$ with that of hardcore bosons.

\section{Quasi-steady state}\label{sec:dynamics}

\subsection{Current\label{sec:steady}}
%After the quench of the hopping phase a finite current is established and a (quasi-)steady state forms in the bulk, similarly to the quasi-steady state current generated in finite 1D fermionic systems (whether interacting or not)~\cite{Chih-Chun:2012}. In fact the current extracted from our simulations has a ballistic character in the superfluid state and does not decay for times $\lesssim 20 \div 200\, t_0$ depending on the filling. In contrast, there is no steady state in the Mott insulator. 
In Fig.~\ref{fig:two} the time evolution of the current in the middle of the chain  is shown in the case of both small (Fig.~\ref{fig:two}\,\textbf{a}-\textbf{b}-\textbf{c}) and large phase quenches (Fig.~\ref{fig:two}\,\textbf{d}-\textbf{e}-\textbf{f}) and for the three different fillings $n=0.1,\,0.5,\,1\,$. The current is extracted from the simulations as the expectation value of the current operator
\begin{equation}\label{eq:current_exp}
\begin{split}
j = \langle {\hat j}\rangle = \frac{iJ}{\hbar}\langle (\hat b_{L/2+1}^\dagger\hat b_{L/2} - \hat b_{L/2}^\dagger\hat b_{L/2+1})\rangle\,,
\end{split}
\end{equation}
and is positive if the particles move from the right to the left of the lattice. Immediately after the quench the current reaches a constant value with negligible fluctuations and this corresponds to the formation of a quasi-steady state. Large oscillations set in after a time that depends on the system size and interaction strength. We emphasize that the formation of a quasi-steady state is not restricted to the low filling limit with emergent Galilean invariance, or to the integrable limit of hardcore bosons (see Section~\ref{sec:noninteracting}), but is a generic feature of the superfluid state of interacting bosons in 1D lattices. It is in fact the signature of a nonzero Drude weight (Section~\ref{sec:drude}). In the Mott insulator regime there is no finite steady-state current, as expected.

The large oscillations at later times shown in Fig.~\ref{fig:two} are a manifestation of the complex dynamics in a finite system with boundaries where the time-evolved density is no longer constant and the steady state cannot be maintained. Ultimately, all of the gas is reflected back at the boundary towards which the current is directed, leading to a \textit{current inversion}. The data for long enough times (if available) show a region of an essentially constant current with equal magnitude but opposite sign compared the plateau immediately after the quench.  Thus the gas not only propagates ballistically in the lattice but is also reflected in a perfectly elastic way at the left boundary.
The current inversion occurs earlier for higher values of $U/J$. This is due to the faster propagation of the shock and rarefaction waves with increasing interaction strength (see Sec.~\ref{sec:shock} below).

This qualitative picture is generic for any value of $U/J$ provided the filling is lower than $n = 1$. For $n = 1$ and $U/J\gtrsim 3.4$~\cite{Lauchli:2008} the system is a Mott insulator. Comparing the results in Fig.~\ref{fig:two} for fillings $n = 0.1$ and $n = 0.5$ with integer filling $n =1$ shows that the dynamics in the Mott insulator regime is qualitatively different since the initial current decays to zero, and at even higher values of $U/J$ the current is identically zero from the beginning. 

The time scales that can be explored in the case of a large phase quench are much shorter with respect to the small phase quench due to the faster build up of entanglement during the evolution, which is largest at the left boundary where the shock wave forms. This explains why the data for longer times are available in the upper panels of Fig.~\ref{fig:two} with respect to the lower panels.
The entanglement growth is faster for larger filling factors and is particularly evident in the case of unit filling $n = 1$ (Fig.~\ref{fig:two}\,\textbf{c}-\textbf{f}) where the dynamics in the Mott insulator state yields a fast entanglement growth and it has been possible to simulate the system only for very short times ($t\lesssim 5t_0$ in Fig.~\ref{fig:two}\,\textbf{f}). On the contrary the ground state of gapped systems such as a Mott insulator is better captured by the MPS ansatz than gapless systems such as a superfluid. The entanglement growth is discussed in Sec.~\ref{sec:entanglement}.

%%%%%%%%%%%%%%%%%%
%%%%% FIG. 3 %%%%%
%%%%%%%%%%%%%%%%%%
\begin{figure}
\includegraphics{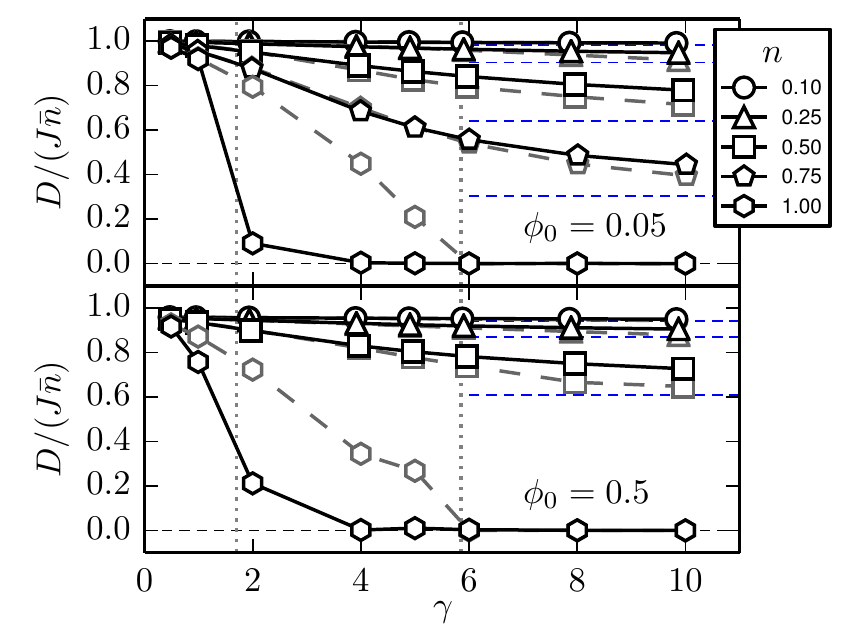}
\caption{\label{fig:three} (Color online) The quantity $D/(J\bar{n})$ as a function of the density-rescaled Hubbard interaction $\gamma = U/(2J\bar{n})$ (or Lieb-Liniger parameter, see Sec.~\ref{sec:noninteracting}). The different symbols correspond to different fillings. The black lines are TDMRG results and the grey lines refer to the time-dependent Gutzwiller ansatz~\cite{Peotta:2013}. The two vertical dotted lines are the exact ($U_{c,{\rm exact}}/J = 3.4$)~\cite{Lauchli:2008} and mean field ($U_{c,{\rm mean\,field}}/J = 11.7$)~\cite{Fisher:1989} values of the critical interaction strength. The horizontal dashed lines are the asymptotic limits in the corresponding hardcore case ($U/J = +\infty$). These data have been extracted from the results for $j(t)$ shown in Fig.~\ref{fig:two} according to Eq.~(\ref{eq:current-carrying_mass_fraction}). The value of the velocity in Eq.~(\ref{eq:current-carrying_mass_fraction}) is taken at time $t^* = 10t_0$ if the data are available. Otherwise $t^*$ is the maximum time reached in each simulation. The upper panel refers to the small phase quench $\phi_0 = 0.05$ while the lower to the large phase quench $\phi_0 = 0.5$.}
\end{figure}

%%%%%%%%%%%%%%%%%%
%%%%% FIG. 4 %%%%%
%%%%%%%%%%%%%%%%%%
\begin{figure}
\includegraphics{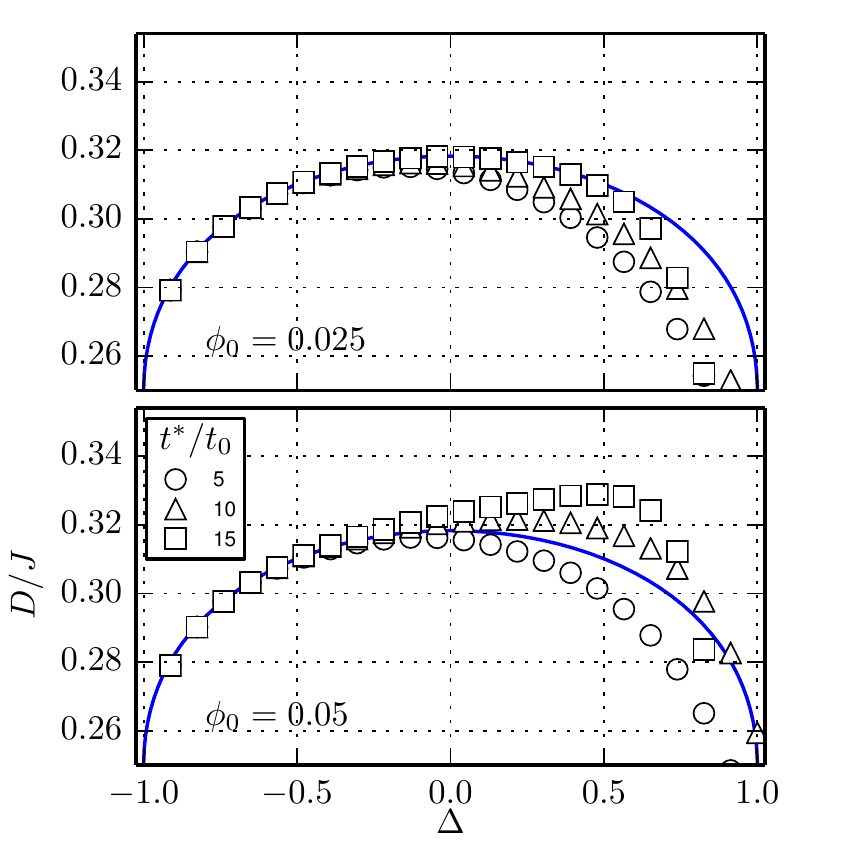}
\caption{\label{fig:four} (Color online)  The symbols shown are the Drude weight extracted from our TDMRG simulations of a phase quench in the XXZ model~(\ref{eq:XXZ}) at half filling for values of the anisotropy parameter in the range  $-1 < \Delta <1$. Eq.~(\ref{eq:current-carrying_mass_fraction}) is used to relate the current obtained from the simulations to the Drude weight. The blue curve is the result of the exact Bethe ansatz solution of the XXZ model~\cite{Shastry:1990} [Eq.~(\ref{eq:drude_exact})]. The upper panel refers to a phase quench with $\phi_0 = 0.025$ while the lower panel to $\phi_0 = 0.05$. Different symbols refer to the time $t^*$ at which the Drude weight is extracted. Note that for $\Delta > 0$ the measured value of $D$ depends both on $t^*$ and $\phi_0$, while this not the case for $\Delta<0$ where a well defined value is obtained which coincides with the Bethe ansatz result.}
\end{figure}

The value of the post-quench hopping phase $\phi_0 = 0.05$ is representative for a kind of quench that does not change the total energy of the system substantially. In other words, the effective temperature after the quench is not far from the temperature before, which is zero in our case. Indeed, if we consider a weakly interacting system with density-rescaled interaction strength $\gamma = U/(2J\bar{n})< 1$ (see Sec.~\ref{sec:noninteracting}), the variation of kinetic energy density due to the quench is $J\bar{n}\phi_0^2$, while the total energy density can be approximated by the potential energy density alone given by $U\bar{n}^2/2$. The density $\bar{n} =\langle \hat{n}_{L/2}+\hat{n}_{L/2+1}\rangle/2$ in the middle of the chain has been used, which can be slightly different from the filling factor $n$ since the density profile at the boundary depends on the interaction strength $U/J$.
Comparing the two energies leads to the inequality $(\phi_0/\bar {n})^2 < \gamma$ that has to be satisfied for the post quench effective temperature to be considered close to zero. This is the case for all values of the parameters used in the simulations. For $\gamma > 1$ the above inequality is a conservative estimate since the total energy of the system increases with $\gamma$ while the variation of kinetic energy due to the quench decreases.
For $\phi_0 = 0.5$ the above inequality is not satisfied with our choice of parameters and thus the system is quenched to a high effective temperature.

Fig.~\ref{fig:two} shows that the current has a similar behavior in the small and large phase quenches (see also Sec.~\ref{sec:drude} and Fig.~\ref{fig:three} below). In the large phase quench case it is more evident that noninteracting bosons do not attain a steady state. In stark contrast, a quasi steady-state current forms even for very weakly interacting bosons. This sharp transition in transport behavior is a clear manifestation of the phase transition from a gapless to a ferromagnetic state in the XXZ model~(\ref{eq:XXZ}) and has been discussed in Ref.~\cite{Chih-Chun:2012}. 

\subsection{Drude weight\label{sec:drude}}
The quench in the hopping phase that we consider here is in fact a simple way to measure experimentally the Drude weight~\cite{Kohn:1964}, which is the strength of the peak at zero frequency of the real part of the conductivity: $\sigma(\omega) = D\delta(\omega) + \sigma_{\rm reg}$, where $D$ is the Drude peak and $\sigma_{\rm reg}$ is the
regular part describing scattering processes at finite energy. The Drude peak effectively measures the amount of dissipationless current that a system can sustain. A convenient way to extract the Drude weight is to calculate the change in the ground-state energy in the presence of an external flux~\cite{Shastry:1990,Giamarchi:1995,Rossini:2011a,Rossini:2011b,Rossini:2013} $\Phi = \oint \mathbf{A}\cdot d\mathbf{l} = L\phi$ when periodic boundary conditions are assumed for the system:
\begin{equation}\label{eq:drude}
D = \frac{L}{2}\left.\frac{d^2E_0}{d\Phi^2}\right|_{\Phi=0} = \frac{1}{2}\left.\frac{d^2E_0/L}{d\phi^2}\right|_{\phi=0}\,.
\end{equation}
On the other hand, the total persistent current $I = Nv$ is given by
\begin{equation}\label{eq:current}
I = \frac{L}{\hbar}\frac{dE_0}{d\Phi}= \frac{L}{\hbar}\frac{dE_0/L}{d\phi}\,,
\end{equation}
Eqs.~(\ref{eq:drude}) and~(\ref{eq:current}) are exact relations between global quantities such as the total ground state energy, the current and the flux. However, we expect them to be valid in the local form shown in the respective right hand sides. The following simple relation follows
\begin{equation}\label{eq:current-carrying_mass_fraction}
\frac{vt_0}{2\phi_0} = \frac{D}{J\bar{n}}\,.
\end{equation}
We have used the velocity $v(t) = j(t)/\bar{n}$ and the density $\bar{n} =\langle \hat{n}_{L/2}+\hat{n}_{L/2+1}\rangle/2$ measured in the middle of the chain which may differ from the filling $n$ due to boundary effects.
Thus the ratio $D/(J\bar{n})$ is the mass fraction that carries the persistent current and can be inferred from our simulations using Eq.~(\ref{eq:current-carrying_mass_fraction}). 

The value of this quantity extracted from our simulations is shown in Fig.~\ref{fig:three}.
To better compare the data for different fillings we show $D/(J\bar{n})$ as a function of the density-rescaled parameter $\gamma = U/(2J{\bar n})$ (see Sec.~\ref{sec:noninteracting}).
The velocity in Eq.~(\ref{eq:current-carrying_mass_fraction}) is measured at $t = 10t_0$ in the case of the small phase quench (upper panel), a time long enough for the initial transient effects to have faded away in most cases (with the exception of the Mott states for $U/J = 4.0$ [$\gamma \sim 2$] which shows a long relaxation time). In the case of the large phase quench (lower panel) $t^*$ is the maximum time reached in each simulation if data at $t = 10t_0$ are not available.
In the weakly interacting limit, which corresponds to the continuum limit~\cite{Delande:2014}, the emergent Galilean invariance of the system fixes the Drude weight to be equal to the total density ($D/J\sim \bar{n}$) as it can be seen in Fig.~\ref{fig:three}.
When the filling is increased the Drude weight decays with increasing $U/J$, eventually leading to an insulating state at $n = 1$ and $U/J > 3.4$, a consequence of the lattice-induced backscattering. We can see that the Gutzwiller ansatz slightly underestimates the steady state velocity for all fillings $n < 1$, but it greatly overestimates it at filling $n=1$ due to the mismatch between the exact ($U_{c,{\rm exact}}/J = 3.4$) and mean field ($U_{c,{\rm mean\,field}}/J = 11.7$)~\cite{Fisher:1989} critical values of the interaction strength.
An important message of Fig.~\ref{fig:three} is that the Drude weight measured by means of a phase quench does not depend in a substantial way on the value of $\phi_0$.

The Drude weight for the Bose-Hubbard model is not known analytically and can be only extracted by numerical methods such as DMRG. In order to validate the scheme employed here to calculate numerically the Drude weight we show in Fig.~\ref{fig:four} results for of the XXZ model~(\ref{eq:XXZ}) at half filling. At half filling the value of the Drude weight is exactly known~\cite{Shastry:1990} and reads
\begin{equation}\label{eq:drude_exact}
\frac{D}{J} = \frac{\pi}{4}\frac{\sin\mu}{\mu(\pi-\mu)}\,,
\end{equation}
with $\cos\mu =\Delta$ (the Drude weight has unit of energy in our case).
 One may expect that the Drude weight, as a ground state property of a system with periodic boundary conditions, cannot be related to the non-equilibrium dynamics of a system with open boundary conditions. The results in Fig.~\ref{fig:four} show that in fact for $-1 < \Delta <0$ (relevant for the Bose-Hubbard model, see Eq.~(\ref{eq:anisotropy_parameter})) the Drude weight can be very precisely extracted from a phase quench as the one considered for the Bose-Hubbard model. On the antiferromagnetic side $0< \Delta <1$ the current relaxes to the equilibrium value on a longer time scale and also finite size effects are more prominent. This can be seen from the fact that the measured value of $D$ depends both on the time $t^*$ at which the steady-state current is taken and on the magnitude of the phase quench $\phi_0$. This behavior is not present on the ferromagnetic side. To avoid unnecessary distraction from the main topic, we will not discuss this finite size effect further, but it is possible that larger system sizes should allow one to extract the Drude weight even for $\Delta > 0$.

In general, the Drude weight is distinct from the superfluid fraction in spatial dimensions lower than three as discussed in Refs.~\cite{Giamarchi:1995,Svistunov:2000,Cazalilla:2011} (see also Ref.~\cite{Scalapino:1992}) due to the fact that the thermodynamic ($L\to +\infty$) limit and zero temperature limit do not commute. This is nicely illustrated by the hardcore (free fermions) case where the superfluid fraction is necessarily zero, but the Drude weight is finite as shown by the horizontal asymptotes in Fig.~\ref{fig:two}. It is an interesting open question if the setup proposed here can be used to measure the superfluid fraction, that coincides with the Drude weight in higher dimension at zero temperature, without the need of rotating the gas (as in the ultracold gas analog of the classical Andronikashvili experiment~\cite{Cooper:2010}). Here we focus on the zero temperature case and address how the Drude weight may be measured in finite systems, however the Drude weight at finite temperature in integrable and nonintegrable 1D spin chains has been studied as well using TDMRG~\cite{Karrasch:2012,Karrasch:2013}.

\subsection{Entanglement entropy dynamics\label{sec:entanglement}}
The entanglement entropy is a crucial quantity relevant to the performance of the TDMRG algorithm~\cite{tdmrg}. It also provides information of correlations in many-body systems. If $\bm{\rho}_i$ is the reduced density matrix obtained by tracing out the states on lattice sites $i+1$ to $L$, then the entanglement entropy is defined as
\begin{equation}
S_i = -\mathrm{Tr}[\bm{\rho}_i\ln \bm{\rho}_i]\,.
\end{equation}
We found that during the evolution the entanglement grows faster as we approach the left boundary of the system, where the shock wave forms, and this is the main reason the time scale reached by TDMRG simulations is limited. Here we present results for the behavior of entanglement in the quasi-steady state away from the boundaries. In particular, we focus on the entanglement entropy between the two halves of the system, $S_{L/2}(t)$. 

It has been noted in Ref.~\cite{KlichLevitov} that in the thermodynamics limit the entanglement entropy production rate is a constant when two noninteracting fermion systems are connected, and
Ref.~\cite{Chien:2014} shows that a quasi-steady state of noninteracting fermions is also characterized by a constant entropy production rate. For noninteracting fermions the rate estimated from the semiclassical full counting statstics \cite{KlichLevitov} is a function of the transmission coefficient of the junction through which the current flows:
\begin{equation}\label{eq:ent_rate}
\frac{dS}{dt} = -\frac{\Delta\mu}{h}\left[ T\log_2 T + (1-T)\log_2 (1-T)\right]\,.
\end{equation}
Here $\Delta\mu$ is the chemical potential difference (whose role is played by $\phi_0$ here) and $T$ is the transmission coefficient.

In our simulations of interacting bosons we found that $dS/dt = 0$ in the superfluid phase. Interestingly, this agrees with Eq.~(\ref{eq:ent_rate}) and the full quantum-mechanical simulations in Ref.~\cite{Chih-Chun:2013} for noninteracting fermions. Since we consider a uniform lattice without any constriction in the middle junction, the transmission coefficient is $1$ so noninteracting fermions do not produce further entanglement entropy. A similar reason may account for the behavior of interacting bosons in the superfluid phase.

\begin{figure}
\includegraphics{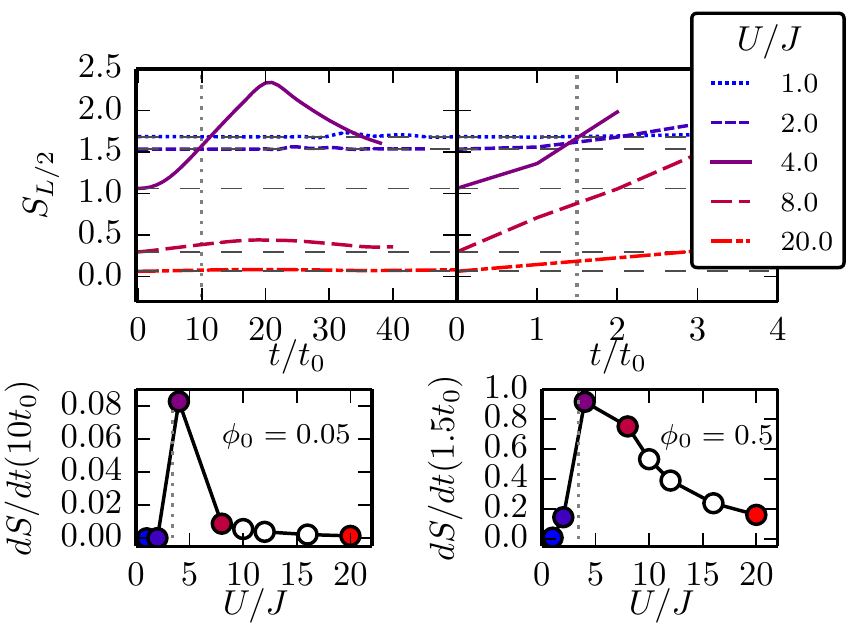}
\caption{\label{fig:five}
(Color online) In the upper panels we show the entropy as a function of time $S_{L/2}(t)$ relative to the bipartition of the system in two halves. The plots from top to bottom correspond to increasing interaction strength ($S_{L/2}(0)$ decreases with increasing $U/J$). On the left results for small phase quench $\phi_0 = 0.05$ are shown and on the right for the large phase quench $\phi_0 = 0.5$. In the lower plots the corresponding values of the entropy production rate $dS_{L/2}(t)/dt$ (in units of $t_0^{-1}$) at time $t = 10t_0$ (left) and $t = 1.5t_0$ (right) are reported as a function of the interaction strength. The filled circles correspond to the values of $U/J$ shown in the upper plots. The vertical dotted lines indicate the critical value of the interaction strength $U_{c, \text{exact}}/J = 3.4$. All of the data refer to filling $n=1$.}
\end{figure}

In contrast, a finite entropy production rate is present around the superfluid-Mott insulator transition at $n = 1$ as shown in Fig.~\ref{fig:five}. Although the entanglement entropy of the initial ground state at $t=0$ is a decreasing function of $U/J$, during the dynamics the entanglement entropy production rate reaches a maximum around the critical point. In fact we see from the lower panels of Fig.~\ref{fig:five} that the rate jumps from essentially zero below the critical point to a finite value right above the critical point. This indicates that the correlation between the two parts of the system increases in the Mott insulator phase close to the critical point after a quench in the external gauge field. This feature of the entanglement entropy may serve as another sharp indicator of the superfluid-Mott insulator transition.

\section{Shock and rarefaction waves\label{sec:shock}}

In the previous Section we have analyzed the current, the Drude weight, and the entanglement entropy that are properties characterizing the bulk of the system. We now concentrate on the time evolution of the density profile $n_i = \langle \hat{n}_i\rangle$ after the quench, especially near to the boundary. The case of a small phase quench $\phi_0 = 0.05$ is discussed first. 

\begin{figure*}
\includegraphics{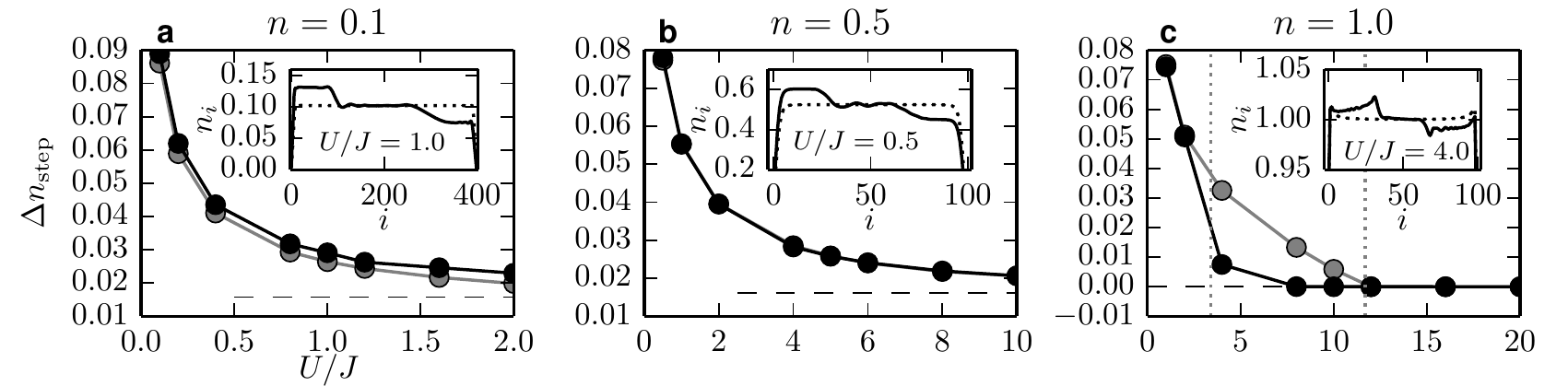}
\caption{\label{fig:six}\textbf{Panels} \textbf{a}-\textbf{b}-\textbf{c}: density step height $\Delta n_{\rm step}$ as a function of the interaction strength at fillings $n = 0.1$ (panel \textbf{a}), $0.5$ (\textbf{b}) and $1$ (\textbf{c}) extracted from the time-dependent Gutzwiller ansatz (grey dots) and TDMRG  (black dots) simulations in the case of a small phase quench $\phi_0 = 0.05$. The step height for increasing interaction tends asymptotically to the  value for hardcore bosons (horizontal dashed line). In Fig.~\ref{fig:five}\textbf{c} the two vertical dotted lines are the exact ($U_{c,{\rm exact}}/J = 3.4$)~\cite{Lauchli:2008} and mean field ($U_{c,{\rm mean\,field}}/J = 11.7$)~\cite{Fisher:1989} values of the critical interaction strength. In the insets a snapshot of the density profile $n_i = \langle \hat{n}_i \rangle$ obtained from TDMRG at $t = 0$ (dashed line) and at a later time [solid line, $t = 266\,t_0$ (panel \textbf{a}), $40\,t_0$ (\textbf{b}), $13\,t_0$ (\textbf{c})] for the selected values of the interaction strength.   The dynamics in a Mott insulator are substantially different  without well-defined shock and rarefaction waves (compare inset of panel \textbf{c} for $U/J = 4.0$ and $n = 1$ to the other insets and Fig.~\ref{fig:one_1}\textbf{b}).}
\end{figure*}

The density profile dynamics is qualitatively similar everywhere in the superfluid region, i.e., away from the Mott insulator state that occurs for $U/J\gtrsim 3.4$ and commensurate filling $n = 1$, and  it is illustrated in Fig.~\ref{fig:one_1}\textbf{b} and in the insets of Fig.~\ref{fig:six}. At the boundaries two density steps form with magnitude $\Delta n_{\rm step}$, where this latter quantity is defined in the caption of Fig.~\ref{fig:one_1}\textbf{b}. For a current flowing toward the left of the system the density step on the left is positive and is connected to the constant background in the middle by a \textit{shock wave}~\citep{Whitham_book} which approximately retains its shape as it propagates in the direction opposite to the bulk current. On the right side the density step is negative and is connected to the background by a region where the slope of the density profile decreases in time (at low filling), \textit{i.e.}, a \textit{rarefaction wave}~\citep{Whitham_book}.
Shock and rarefaction waves play an important role in the dynamics of ultracold gases, but despite much effort they are still poorly understood (see, e.g., Ref.~\cite{Dutton:2001,Hoefer:2006,Meppelink:2009,Joseph:2011,Bulgac:2012,Salasnich:2011,
Ancilotto:2012,Bettelheim:2012,Mirlin:2012,Kulkarni:2012,Lowman:2013,Peotta:2014a} and references therein).

The definition of the step height $\Delta n_{\rm step} = n_i(t)-\bar{n}$ provided in Fig.~\ref{fig:one_1} is relatively insensitive to the time $t$ and the lattice site $i$ at which it is evaluated provided that (i) $i$ is a site in between (and sufficiently far from both) the boundary and the shock front, and (ii) at time $t$ the front has traveled a long enough distance.
The only exception is right above the transition where no well-defined plateau appears. The value of $\Delta n_{\rm step}$ reported for $U/J = 4.0$ in Fig.~\ref{fig:six}\,\textbf{c} gives an estimate of the magnitude of the density perturbation induced by the quench, as shown in the inset, but this is not the height of a well-defined density step as for all other points.

The absence of shock and rarefaction waves is a sensitive dynamical probe of the Mott insulator which can be measured in experiments with uniform gases~\cite{Hadzibabic:2013,Hadzibabic:2014,Hadzibabic:2014-2}. The density profile is measurable in experiments and the current can be inferred from it at different times using the protocol outlined in Ref.~\cite{Chih-Chun:2012}.

The Gutzwiller ansatz is capable of capturing the step height quite well as shown in Fig.~\ref{fig:six}, except at the transition between the superfluid state and the Mott insulator. This discrepancy is similar to the one observed in Fig.~\ref{fig:three}. It appears that above the transition to the Mott insulator state the correlations, which are not well captured by the mean-field Gutzwiller ansatz, are crucial in obtaining the correct dynamics.

\begin{figure}
\includegraphics[scale=1.0]{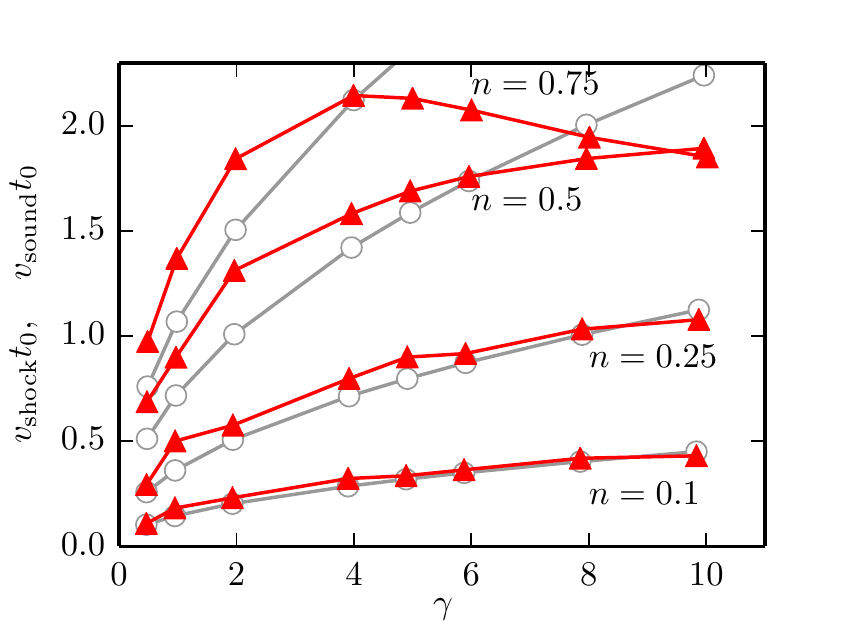}
\caption{\label{fig:seven} (Color online) The shock wave propagation speed $v_{\rm shock} = \bar{n} v/\Delta n_{\rm step}$ (triangles) for different initial fillings ($n = 0.1, 0.25, 0.5, 0.75$ from bottom to top) compared to the sound speed in a weakly interacting Bose gas $v_{\rm sound}t_0 = 2\bar{n}\sqrt{\gamma}$ (circles) as a function of the density-rescaled interaction strength $\gamma = U/(2J\bar{n})$.}
\end{figure}

The density step height $\Delta n_\text{step}$ can be used to estimate by current conservation the velocity of propagation of the front $v_{\rm shock}$ (see Fig.~\ref{fig:one_1}\,\textbf{b})  according to the formula $v_{\rm shock} = \bar{n}v/\Delta n_{\rm step} = j/\Delta n_{\rm step}$ where the bulk velocity $v$ and bulk current $j$ have been considered in Sec.~\ref{sec:steady} and~\ref{sec:drude}. In Fig.~\ref{fig:seven} we show the propagation speed of the shock waves at the left as a function of the parameter $\gamma = U/(2J\bar{n})$. The results of the numerical simulations are compared with the sound velocity for a weakly interacting Bose gas $v_{\rm sound}(\gamma)t_0 = \sqrt{2\bar{n}U/J} = 2\bar{n}\sqrt{\gamma}$~\cite{Cazalilla:2011}.

The shock propagates at a speed which is very close to the sound one at low filling. The sound speed at higher filling still provides a good order of magnitude estimate of the shock wave propagation speed. For low values of $\gamma$ the relation  $v_{\rm shock}(\gamma) > v_{\rm sound}(\gamma)$ holds in general, but eventually the opposite inequality takes place for large enough $\gamma$. In the case $n = 0.75$ the slope of $v_{\rm shock}$ becomes negative for $\gamma > 4$. 
%\textbf{This behavior is a consequence of the fact the near to the hardcore lim%it $U/J\to +\infty$ the mass transport can be equivalently understood in terms %of holes (unoccupied states) rather than particles, and the density of holes de%creases with increasing filling. (This a very hand waiving explanation and I am% not sure it is correct)}
 A qualitative explanation of why the shock wave speed is close or slightly larger than the sound speed, at least for small interaction strength, remains a challenge. Only in the low filling (continuum limit) the propagation speed and the density step of the shock wave are controlled by the dispersive analogue~\cite{Hoefer:2013} of the Hugoniot loci for classical shock waves~\cite{Whitham_book} since viscosity is extremely low in ultracold gases.
 
Whereas in the case of the current, Drude weight, and entanglement entropy the results obtained with the small and large phase quenches are quite similar, this is not the case for the time evolution of the density profile at the boundaries. No well defined density steps and propagating wavefronts have been observed for $\phi_0= 0.5$. The main difference is that the \textit{shock structure}~\cite{Whitham_book} (i.e. the region connecting the constant density plateaus) is more broad in this case and exhibit a more pronounced oscillatory structure typical of dispersive shock waves~\cite{Hoefer:2013}. In fact this is in agreement with the solution of the piston problem for a dispersive fluid~\cite{Hoefer:2008}. It is predicted that for large enough piston velocity (analogous to the post-quench hopping phase $\phi_0$ in our case) a bifurcation of the dynamics occurs and instead of a constant density plateau a locally periodic wave train is generated. Therefore $\Delta n_\text{step}$ and $v_\text{shock}$ can not be defined. This is another indication that a superfluid is in fact a dispersive inviscid fluid.

\section{Conclusion}\label{sec:conclusion}
This work presents our studies of quasi-steady states of interacting bosons in a 1D optical lattice after a quench of an artificial gauge field. We studied the full crossover between the low filling (continuum limit) up to commensurate filling $n=1$ where a Mott insulator forms for strong enough interaction strength.
In the superfluid state we find that once a finite momentum transfer is delivered to the system, shock and rarefaction waves form at the hard-wall boundaries that break the lattice translational invariance. We characterized the shock waves by the density step height and the speed of the ballistic propagation of the wave front. The dynamics is rather different in the Mott insulator as the current is suppressed. The absence of well defined shock and rarefaction wave is a readily measurable dynamical feature of the Mott insulator. Another interesting dynamical property is that at the critical point between the Mott insulator and the superfluid the entanglement entropy production rate is maximal.

We present data for the bulk current of interacting Bose gases and study the lattice-induced correlations that lead to its decay with increasing filling and interaction strength. The current is found to be proportional to the Drude weight of the system and thereby we establish a possible experimental probe of this quantity by using the time-dependent density profiles as an input. 
A question for future research is if the quasi-steady state current after a quench is related to the superfluid fraction in the case of nonzero temperature and higher dimensionality.
An affirmative answer to this question would allow to measure in an exceedingly simple way the superfluid fraction of an ultracold atomic gas, a very important property.

We have considered also a quench with a large value of the post-quench hopping phase.
While bulk properties such as current, Drude weight and entanglement entropy are essentially the same for small and large phase quenches, the behavior at the boundaries is very different in agreement with recent theoretical predictions~\cite{Hoefer:2008}. In the case of a large phase quench it is possible to observe a sharp transition in the formation of a quasi-steady state away from the noninteracting-boson limit $U = 0$ which serves as a dynamical signature for the phase transition between noninteracting and interacting bosons. Finally, the existence of a quasi-steady state current in interacting bosons paves the way for studying interesting transport phenomena in bosonic systems.

\begin{acknowledgments}
This work has been supported by DOE under Grant No. DE-FG02-05ER46204. S. P. acknowledges useful discussions with Nicholas K. Lowman. The numerical results presented in this work have been obtained by using the TDMRG code developed by S.P. in collaboration with Davide Rossini.
\end{acknowledgments}

\bibliographystyle{apsrev4-1}

\end{document}